\documentclass[11pt]{article}
\pdfoutput=1
\usepackage[utf8]{inputenc}
\usepackage{verbatim}
\usepackage{amsmath,amssymb,array}
\usepackage{amsthm}
\usepackage{float}
\usepackage[algoruled,linesnumbered,noend]{algorithm2e}
\usepackage{ifdraft}
\usepackage{cite}
\usepackage{rotating}
\usepackage{multirow}
\usepackage{verbatim}
\usepackage{url}
\usepackage{xspace}
\usepackage{graphicx}
\newcommand{\coll}[1]{\url{#1}}

\usepackage{verbatim}
\newcommand{\nota}[1]{%
  {\sffamily\bfseries #1}%
  \marginpar{\framebox{\Large *}}%
}
\newcommand{\notaestesa}[2]{%
  {\sffamily {\bfseries #1}{\footnotesize #2}}%
  \marginpar{\framebox{\Large *}}%
}

\usepackage{tikz}
\usetikzlibrary{shapes,arrows}
\usetikzlibrary{snakes}
\usetikzlibrary{positioning,patterns}
\usepackage{subfigure}

\usepackage{rotating}

\usepackage{ifpdf}\usepackage{datetime}
\ifpdf
\fi

\usepackage{booktabs}

\usepackage{tabularx}
\newcolumntype{T}[1]{>{\tsize} #1}
\newcolumntype{W}{>{\raggedleft\arraybackslash}X}
\newcolumntype{C}{>{\centering\arraybackslash}X}
\usepackage{multirow}
\usepackage{longtable}
\usepackage{array}

\newcommand{\tsize}{\scriptsize}

\newcommand{\ie}{i.e.~}
\newcommand{\st}{s.t.~}
\newcommand{\wrt}{w.r.t.~}
\newcommand{\ExGene}{\ensuremath{\langle B,F\rangle}}

\DeclareMathOperator{\prefix}{pre}
\DeclareMathOperator{\suffix}{suf}
\DeclareMathOperator{\enc}{enc}
\DeclareMathOperator{\FH}{LH}
\DeclareMathOperator{\SH}{RH}
\newcommand{\Ll}{\ensuremath{\mathcal{L}_l}\xspace}
\newcommand{\Lr}{\ensuremath{\mathcal{L}_r}\xspace}

\renewcommand{\emptyset}{\ensuremath{\varnothing}}
\newcommand{\GS}{\ensuremath{G_{I}}}

\begin{document}

\newtheorem{theorem}{theorem}
\newtheorem{lemma}[theorem]{Lemma}
\newtheorem{proposition}[theorem]{Proposition}
\newtheorem{observation}[theorem]{Observation}
\newtheorem{corollary}[theorem]{Corollary}
\newtheorem{claim}[theorem]{Claim}
\newtheorem{property}[theorem]{Property}
\theoremstyle{definition}
\newtheorem{definition}[theorem]{Definition}
\newtheorem{problem}{Problem}
\newtheorem{example}{example}

\title{Reconstructing Isoform  Graphs  from RNA-Seq data
}

   \author{Stefano Beretta, Paola Bonizzoni, Gianluca Della Vedova,
       Raffaella Rizzi\\
Univ. Milano-Bicocca, 
Milano, Italy\\
\url{http://algolab.eu}}

\maketitle

\begin{abstract}
Next-generation sequencing (NGS)
technologies allow new methodologies for  alternative splicing  (AS)
analysis. Current computational methods for AS  from NGS data are mainly
focused on predicting splice site junctions or  de novo assembly of
full-length transcripts. These methods are computationally expensive and
produce a huge number of  full-length transcripts or splice junctions,
spanning the whole genome of organisms.  Thus summarizing such data  into the different gene structures
 and AS events of the expressed genes is an hard task.

To face this issue in  this paper we investigate the computational
problem of reconstructing from NGS data, in absence of the genome,   a gene
structure for each gene that is represented by the \emph{isoform graph}: we
introduce such graph and we show that it  uniquely   summarizes the
gene transcripts.
We define the computational problem of reconstructing the isoform graph and
provide some conditions that must be met to allow such reconstruction.
Finally, we describe an efficient algorithmic  approach to solve this problem,
validating our approach with both   a theoretical and an  experimental analysis.
\end{abstract}

\section{Introduction}
\label{sec:intro}

Next-generation sequencing (NGS) technologies allow massive and
parallel sequencing of biological molecules (DNA and RNA), and have a
huge impact on molecular biology and
bioinformatics~\cite{Metzker2010}. In particular, RNA-Seq is a recent
technique to sequence expressed transcripts, characterizing both the
type and the quantity of transcripts expressed in a cell (its
transcriptome).  Challenging tasks of transcriptome analysis via
RNA-Seq data analysis ~\cite{Trapnell2010,Nicolae2011,FengTao} are
reconstructing full-length transcripts (or isoforms) of genes and
their expression levels.  The most recent   studies indicate that alternative
splicing is a major mechanism generating functional diversity in
humans and vertebrates, as at least 90\% of human genes exhibit splicing variants.
The annotation of alternative splicing
variants  and AS events, to differentiate and compare organisms, is part of the
central goal in transcriptome analysis of identifying and
quantifying all full-length transcripts.  However,  the extraction  of splicing variants or significant AS events
from the different transcripts produced by a set of genes requires to compare hundreds of
thousands of full-length transcripts.  Graph representations of splice
variants, such as splicing graphs~\cite{HeberASTP02}, have emerged as
a convenient approach to summarize several transcripts from a gene
into the putative gene structure they support.  The current notions of splicing graphs rely on some sort
of gene annotations, such as the annotation  of full-length transcripts by their constitutive exons on the genome.

In this paper, we first define the notion of \emph{isoform graph} which is a gene structure representation of
genome annotated full-length transcripts of a gene. The isoform graph is a variant of the notion of splicing graph  that has been originally introduced in~\cite{wabi/LacroixSGB08} in a slightly different setting.
When using only RNA-Seq data, the genome annotation  cannot be given, and thus it is
quite natural to investigate and characterize  the notion of  splicing graph  which naturally arises
when  a reference genome is not known or available.  Thus, in the paper we focus on the following main question:
\emph{under which  conditions  the reconstruction of a gene structure can be efficiently accomplished
using only information provided by RNA-Seq data?}

In order to face this problem,  we give some necessary or sufficient
conditions to infer the isoform graph, we introduce an optimization problem
that guides towards finding a good approximation of the  isoform graph and
finally we describe an efficient heuristic for our problem on data violating
the conditions necessary to exactly infer the isoform graph.

The novelty of our approach  relies on the fact that it allows the reconstruction of
the splicing graph    in absence of the genome, and thus it is applicable
also to
highly
fragmented or altered data, as found in cancer genomes.
Moreover we focus on methods that can effectively be used for a genome-wide
analysis on a standard workstation.


Our computational approach to AS  is different from  current methods of transcriptome analysis that focus on using
RNA-Seq data for reconstructing the set of transcripts (or isoforms)
expressed by a gene, and estimating their abundance.  In fact, we aim on
summarizing genome-wide RNA-Seq data  into graphs
each  representing an   expressed gene  and the alternative splicing events
occurring in the  specific processed sample. On the contrary, current tools
do not give a concise result, such as a structure for each gene, nor they
provide a  easy-to-understand listing of AS events for a gene.

Observe  that in absence of a reference, the information given by the
predicted full-length transcripts spanning the whole genes does not imply a
direct procedure to built the isoform graph. In fact, the annotation against a
genome is  required to compute such a graph, as we need to determine from
which gene a given  transcript is originated, may lead to a complex and time consuming  procedure.

Our paper is not focused on  full-length transcripts reconstruction~\cite{Garber2011}, such as
Cufflinks~\cite{Trapnell2010}, and Scripture~\cite{Guttman2010} or
de novo assembly methods that build a de Brujin graph such as
TransAbyss~\cite{Robertson2010}, Velvet~\cite{Zerbino2008}, and
Trinity~\cite{Grabherr2011}. These tools are computationally expensive and are
able to find only the majority of the annotated isoforms while providing  a
large amount of non annotated full-length transcripts that would need to be experimentally validated.

On the other end, the use of de Brujin graph to build approximations of splicing graphs
reveals clear shortcomings due to repeated sequences in distinct genes that lead to assembly chimeric transcripts
and hence fusions of RNA-Seq data from distinct gene structures into unique graphs.

In the paper,    we aim to advance towards
the understanding  of the possibilities and limitations of
computing   the distinct gene structures  from which   genome-wide RNA-Seq or short reads  data have been extracted.
In this sense   our   approach aims  to keep   separate gene structures in the reconstruction from
genome wide RNA-Seq even in absence of a reference.

In this paper, we validate our approach from both theoretical and  experimental points of view.
First we will prove that some conditions must be met in order to guarantee
the reconstruction of the isoform graph from RNA-Seq data.
Then we describe a simple and efficient algorithm that reconstructs the
isoform graph under some more restrictive conditions.
At the same time, a more refined algorithm (sharing the main ideas of the
basic one) is able to handle instances where the aforementioned conditions do
not hold due to, for example, errors in the reads or lower coverage that
typically affect real data.

We show experimentally, as well as theoretically, the scalability  of our
implementation to  huge quantity of data.
 In fact limiting the time and space
computational resources used by our algorithm is a main aim of ours,
when compared to other tools of transcriptome analysis.
More precisely, our algorithmic approach works in time  that is linear in the number of reads, while
having space requirements bounded by the size of hashing tables used to
memorize reads.

Moreover, we are able to show that are method is able to distinguish among different gene structures though processing
a set of reads from various genes, limiting the process of fusion of  graphs structures from distinct genes due to  the presence
of repeated sequences.

The theoretical and experimental analysis have pointed out limitations that are inherent the input data.
As such, we plan to further improve our algorithms and our implementations to
be able to  deal with the different situations coming from these limitations.



\section{The Isoform Graph and the SGR Problem}
\label{subsec:isoform_graph}

A conceptual tool that has been used to investigate the reconstruction of full-length transcripts from
ESTs (Expressed Sequence Tags) \cite{HeberASTP02} or RNA-Seq data is the  {\em splicing graph}.
In this paper we use a notion of a splicing graph
that  is close to the one in~\cite{wabi/LacroixSGB08}, where splicing graphs provide a representation of  variants.
Since our main goal is the reconstruction of the splicing graph  from
the nucleotide sequences of a set of short reads without the knowledge of  a genomic sequence,
  some definitions will be slightly different from~\cite{wabi/LacroixSGB08}
  where the notion of abundance of reads spanning
some splicing junction sites is central.
Moreover our goal is to study transcripts data originating from different tissues
or samples, where the expression level of each single transcript greatly varies
among samples.
Therefore introducing abundance into consideration is likely to introduce a
number of complications in the model as well as in the algorithms, while
increasing the presence of possible confounding factors.

Informally, a splicing graph is the graph representation of
a gene structure, inferred from a set of RNA-Seq data, where  isoforms correspond to paths of the splicing graphs, while
splicing events correspond to specific subgraphs.


Let $s=s_1s_2\cdots s_{|s|}$ be a sequence of characters, that is a
\emph{string}. Then $s[i:j]$ denotes the substring $s_is_{i+1}\cdots s_j$ of
$s$, while $s[:i]$ and $s[j:]$ denote respectively the \emph{prefix} of $s$
consisting of $i$ symbols and the \emph{suffix} of $s$ starting with the
$j$-th symbol of $s$. We denote with $\prefix(s,i)$ and $\suffix(s,i)$
respectively the prefix and the suffix of length $i$ of $s$. Among all
prefixes and suffixes, we are especially interested into
$\FH(s)=\prefix(s,|s|/2)$ and $\SH(s)=\suffix(s,|s|/2)$ which are called the
\emph{left half} and the \emph{right half} of $s$.
Given two strings $s_1$ and $s_2$, the \emph{overlap}
$ov(s_1, s_2)$ is the length of the longest suffix of $s_1$ that is
also a prefix of $s_2$.  The \emph{fusion} of $s_1$ and $s_2$, denoted
by $\varphi(s_1,s_2)$, is the string $s_1[:|s_1|-ov(s_1,s_2)]s_2$
obtained by concatenating $s_1$ and $s_2$ after removing from $s_1$
its longest suffix that is also a prefix of $s_2$. We extend the
notion of fusion to a sequence of strings $\langle s_1, \ldots ,
s_k\rangle$ as $\varphi(\langle s_1, \ldots , s_k\rangle)=
\varphi(s_1,\varphi(\langle s_2, \ldots , s_k\rangle))$ if $k>2$, and
$\varphi(\langle s_1, s_2\rangle)=\varphi(s_1, s_2)$.

In this paper we consider discrete genomic regions (i.e. a gene or a set of
genes) and their full-length isoforms or transcript
products of the genes along these regions.
A gene isoform is a concatenation of some of
the coding regions of the gene respecting their order in
the genomic region.
Alternative splicing regulates how different coding regions are included
to produce different full-length isoforms or transcripts, which are modeled here
as sequences of \emph{blocks}.
Formally, a \emph{block} consists of a string, typically
taken over the
alphabet $\Sigma=\{a,c,g,t\}$, and an integer called rank,
such that no two blocks share the same rank.
The purpose of introducing the rank of a block is
to disambiguate between blocks sharing the same nucleotide sequence (i.e. string) and to
give an order among blocks, reproducing the order of exons in the genomic
region.

\begin{figure}[t]
\newcommand{\wid}{1.1 cm}
\newcommand{\widblock}{2.2 cm}
\newcommand{\dist}{1.1 cm}
\newcommand{\distblock}{2 cm}
\tikzstyle{Exon}=[draw, rectangle, minimum height=0.4cm, minimum
width=\wid, fill=white,anchor=south west]
\tikzstyle{Block}=[draw, rectangle, minimum height=0.4cm, minimum
width=\widblock, fill=white,anchor=south west]
\centering
\subfigure[Exon Skipping.]{
\begin{tikzpicture}[x=\dist, y=-0.6cm, node distance=0.2 cm and 0.9 cm,outer sep =
0pt,->,>=stealth',shorten >=1pt]
\node at (0.4,1.25) {Isoforms};
\node[Exon] at (1,1) {A};
\node[Exon] at (2,1) {B};
\node[Exon] at (3,1) {C};
\node[Exon] at (4,1) {D};
\node[Exon] at (1,2) {A};
\node[Exon] at (4,2) {D};
\draw[-,dashed](2,1)--(2,4);
\draw[-,dashed](4,1)--(4,4);

\node at (0.5,3.65) {Blocks};
\node[Exon] at (1,4) {$b_1$};
\node[Block] at (2,4) {$b_2$};
\node[Exon] at (4,4) {$b_3$};

\node at (0.5,6.65) {Graph};
\node (A) at (1.3,6) {$b_1$};
\node (B) [below right=of A] {$b_2$};
\node (C) [above right=of B] {$b_3$};
\path (A) edge node{} (B)
      (B) edge node{} (C)
      (A) edge [bend left] node{} (C);
\end{tikzpicture}
\label{subfig:skipping}
}
\hspace{0.5cm}
\subfigure[Alternative Donor Site and Mutually Exclusive Exons.]{
\begin{tikzpicture}[x=\dist, y=-0.6cm, node distance=0.2cm and 0.44 cm,outer sep =
0pt,->,>=stealth',shorten >=1pt]
\node at (0.4,1) {Isoforms};
\node[Exon, minimum width=0.64cm] at (1,1) {A'};
\node[Exon] at (3,1) {C};
\node[Exon] at (2,2) {B};
\node[Exon] at (2,1) {B};
\node[Exon] at (5,1) {E};
\node[Exon] at (1,2) {A};
\node[Exon] at (4,2) {D};
\node[Exon] at (5,2) {E};

\draw[-,dashed](1.58,1)--(1.58,4.08);
\draw[-,dashed](2,1)--(2,4.08);
\draw[-,dashed](3,1)--(3,4.08);
\draw[-,dashed](4,1)--(4,4.08);
\draw[-,dashed](5,1)--(5,4.08);

\node at (0.5,3.65) {Blocks};
\node[Exon, minimum width=0.7cm] at (1,4) {$b_1$};
\node[Exon, minimum width=0.5cm] at (1.58,4) {$b_2$};
\node[Exon] at (2,4) {$b_3$};
\node[Exon] at (3,4) {$b_4$};
\node[Exon] at (4,4) {$b_5$};
\node[Exon] at (5,4) {$b_6$};

\node at (0.3,6.5) {Graph};
\node (b1) at (1.2,6) {$b_1$};
\node (b2) [below right=of b1] {$b_2$};
\node (b3) [above right=of b1] {$b_3$};
\node (b4) [right=of b2] {$b_4$};
\node (b5) [right=of b3] {$b_5$};
\node (b6) [below right=of b5] {$b_6$};
\path (b1) edge node{} (b2)
      (b1) edge node{} (b3)
      (b2) edge node{} (b3)
      (b3) edge node{} (b4)
      (b4) edge node{} (b6)
      (b5) edge node{} (b6)
      (b3) edge node{} (b5);
\end{tikzpicture}
\label{subfig:mutual}
}
\caption{Examples of Isoform  Graphs. 
  Capital letters correspond to exons.
  In \subref{subfig:skipping} is represented a skipping of the two
  consecutive exons $B$ and $C$ of the second isoform \wrt the
  first one.
%
  In \subref{subfig:mutual}  is represented an alternative donor site variant
  between exons $A$ and $A'$ and two mutually exclusive exons $C$ and $D$.
  Notice that, in this figure, the isoforms represented are classical.
}
\label{fig:esempio}
\end{figure}
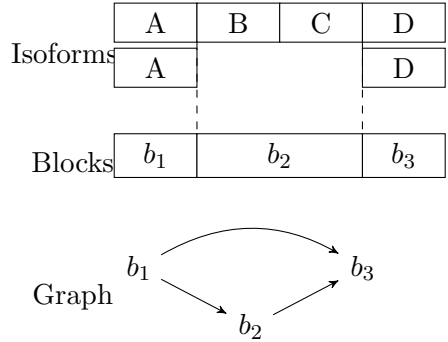
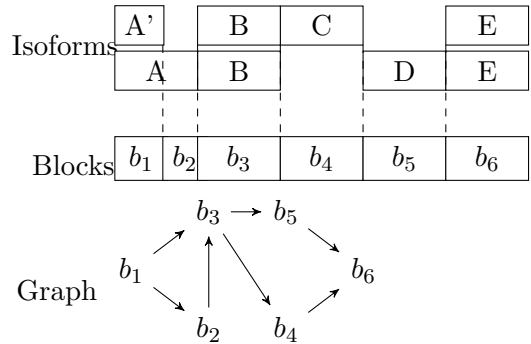

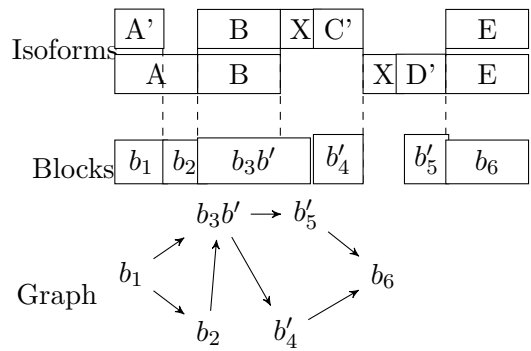
\begin{figure}[t]
\newcommand{\wid}{1.1 cm}

\newcommand{\widblock}{2.2 cm}

\newcommand{\dist}{1.1 cm}
\newcommand{\distblock}{2 cm}
\tikzstyle{Exon}=[draw, rectangle, minimum height=0.4cm, minimum
width=\wid, fill=white,anchor=south west]

\tikzstyle{Block}=[draw, rectangle, minimum height=0.4cm, minimum
width=\widblock, fill=white,anchor=south west]

\centering

\subfigure[Splicing graph compatible with the set of isoforms]{
\begin{tikzpicture}[x=\dist, y=-0.6cm, node distance=0.2cm and 0.44 cm,outer sep =
0pt,->,>=stealth',shorten >=1pt]
\node at (0.4,1) {Isoforms};
\node[Exon, minimum width=0.64cm] at (1,1) {A'};
\node[Exon,minimum width=0.4cm] at (3,1) {X};
\node[Exon,minimum width=0.66cm] at (3.4,1) {C'};
\node[Exon] at (2,2) {B};
\node[Exon] at (2,1) {B};
\node[Exon] at (5,1) {E};
\node[Exon] at (1,2) {A};
\node[Exon,minimum width=0.4cm] at (4,2) {X};
\node[Exon,minimum width=0.66cm] at (4.4,2) {D'};
\node[Exon] at (5,2) {E};

\draw[-,dashed](1.58,1)--(1.58,4.08);
\draw[-,dashed](2,1)--(2,4.08);
\draw[-,dashed](3,1)--(3,4.08);
\draw[-,dashed](4,1)--(4,4.08);
\draw[-,dashed](5,1)--(5,4.08);

\node at (0.5,3.65) {Blocks};
\node[Exon, minimum width=0.7cm] at (1,4) {$b_1$};
\node[Exon, minimum width=0.5cm] at (1.58,4) {$b_2$};
\node[Block,minimum width=1.5cm] at (2,4) {$b_3b'$};
\node[Block,minimum width=0.66cm] at (3.4,4) {$b'_4$};
\node[Exon,minimum width=0.4cm] at (4.5,4) {$b'_5$};
\node[Exon] at (5,4) {$b_6$};

\node at (0.3,6.5) {Graph};
\node (b1) at (1.2,6) {$b_1$};
\node (b2) [below right=of b1] {$b_2$};
\node (b3) [above right=of b1] {$b_3 b'$};
\node (b4) [right=of b2] {$b'_4$};
\node (b5) [right=of b3] {$b'_5$};
\node (b6) [below right=of b5] {$b_6$};
\path (b1) edge node{} (b2)
      (b1) edge node{} (b3)
      (b2) edge node{} (b3)
      (b3) edge node{} (b4)
      (b4) edge node{} (b6)
      (b5) edge node{} (b6)
      (b3) edge node{} (b5);
\end{tikzpicture}
\label{subfig:mutual1}
}
\caption{Alternative splicing graph compatible with the reads of the expressed
  gene of Fig.~\ref{fig:esempio}\subref{subfig:mutual}}
\label{fig:esempio1}
\end{figure}

Given a block
$b$, we denote by $s(b)$ and $r(b)$ its string and rank
respectively.
%
%
In our framework a \emph{gene coding region} is a sequence (that is, an
ordered set) $B=\langle
b_1,b_2,\cdots,b_n \rangle$ of blocks with
$r(b_i)=i$ for each $i$.  Then, the
\emph{string coding region for $B$} is the string $s(b_1) s(b_2)
\cdots s(b_n)$ obtained by orderly concatenating  the strings of the blocks in
$B$.
Intuitively a gene coding region is the sequence of all the coding
regions on the whole genomic sequence for the studied gene.  We define a
\emph{block isoform} $f$
\emph{compatible} with $B$, as a subsequence of $B$, that is $f=\langle
b_{i_1}, \cdots, b_{i_k} \rangle$ where $i_j<i_{j+1}$ for $1\le j< k$.
We distinguish between classical isoforms (defined on exons or
genomic regions) and block isoforms (defined on blocks).
Nonetheless, we will use interchangeably the terms isoforms and block isoforms
whenever no ambiguity arises.
By a slight abuse of language we define the string of $f$, denoted by
$s(f)$, as the concatenation of the strings of the blocks of $f$.


\begin{definition}
\label{expressed}
An \emph{expressed gene} is a pair $\ExGene$, where $B$ is a gene
coding region, $F$ is a set of block isoforms compatible with $B$ where
(1) each block of $B$ appears in some isoform of $F$, and (2)
for each
pair $(b_i, b_{j})$ of  blocks of $B$, appearing
consecutively in some isoform of $F$, there exists a
isoform $f\in F$ \st exactly one of $b_i$ or $b_{j}$ appears in
$f$.
\end{definition}


We point out that Def.~\ref{expressed}  is mostly compatible with that
of~\cite{wabi/LacroixSGB08}, where a \emph{block} is defined as a maximal sequence of
adjacent exons, or exons fragments, that always appear together in a set of isoforms or variants.
Therefore their approach downplays the relevance of blocks with the same string.
Observe that
Def.~\ref{expressed} implies that the set $B$ of blocks of a
string coding region of an expressed region $\ExGene$ is unique and is a
minimum-cardinality set explaining all isoforms in $F$.  Thus, the
pair $\ExGene$ describes a specific gene.

The uniqueness of blocks of an  expressed gene allows us to define the
associated graph representation, or isoform graph.
Given an expressed gene $\mathcal{G}=\ExGene$,  the
\emph{isoform graph} of $\mathcal{G}$ is a directed graph
$G_I=\langle B,E\rangle$, where  an ordered pair $\langle  b_i, b_j
  \rangle$ is an arc of $E$, iff $b_i$ and
$b_j$ are consecutive in at least an isoform of $F$.  Notice that
$G_I$ is a directed acyclic graph, since the sequence $B$ is also a
topological order of $G_I$.
Notice that isoforms correspond to paths in $G_I$.

Our first aim of the paper is to characterize when and how accurately the
isoform graph of an expressed gene can be reconstructed from a set of
substrings (i.e. RNA-Seq data) of the isoforms of the gene.
More precisely, we want to investigate the following two main questions.

\emph{Question 1:} What are the conditions under which the isoform graph of a
gene   can be reconstructed  from a sample of RNA-Seqs (without putting any
bounds on the computational resources)?

\emph{Question 2:} Can we build efficiently such a graph or an approximation of it?

Notice that the isoform graph is the real gene structure that we would like to
infer from data but, at the same time, we must understand that the transcript
data might not be sufficient to determine the isoform graph, as
we have no information on the genomic sequence and on the blocks in particular.
Therefore we aim at computing a slightly less informative kind of graph: the
\emph{splicing graph}, which is a directed graph where each vertex $v$ is labeled
by a string $s(v)$.
Notice that the splicing graph gives no assurance that a vertex is a block, not
does it contain any indication regarding whether (and where) the string
labeling a vertex appears in the genomic region.

For instance, let us consider the isoform graph of Figure~\ref{fig:esempio} (b).
Assume that $s(b_4)$ and $s(b_5)$ share a common
prefix $s(b')$, that is the exons $C$ and $D$ can be respectively written as
$XC'$ and $XD'$.  Then if no information about the block positions and rank on the genomic sequence
is provided as input data, the splicing graph of Figure~\ref{fig:esempio1}   could be as plausible as the
isoform graph of Figure~\ref{fig:esempio} (b).
%
%
%
This  observation leads us to the  notion of splicing graph  \emph{compatible} with a
set of isoforms.

\begin{definition}
\label{def:compatible-graph}
Let $\ExGene$ be an expressed gene, and let  $G=\langle V,E \rangle$ be a
splicing graph.
Then $G$ is \emph{compatible} with $F$ if, for each isoform $f\in
F$, there exists a path $p=\langle w_{1}, \ldots , w_{k}\rangle $ of $G$ such
that $s(f)=s(w_{1})\cdots s(w_{k})$.
\end{definition}

Since we have no information on the blocks of the expressed gene, computing
any graph that is compatible with the isoform graph is an acceptable answer to
our problem.
We need some more definitions related to the fact that we investigate the
problem of reconstructing  a splicing graph compatible with a set of isoforms
only from RNA-Seqs  obtained from the gene
transcripts.
%
 Let $\ExGene$ be an unknown expressed
gene. Then, a \emph{RNA-Seq read} (or simply \emph{read}) extracted
from  $\ExGene$, is a substring of
the string $s(f)$ of some isoform $f \in F$.
Notice that we know only the nucleotide sequence of each read.
Just as we have introduced the notion of splicing graph compatible with a set of
isoforms, we can define the notion of splicing graph compatible with a set of reads.

\begin{definition}
\label{def:read-compatible-graph}
Let  $R$ be a set of reads extracted from an expressed gene $\ExGene$, and let
$G=\langle V,E \rangle$ be a splicing graph.
Then $G$ is \emph{compatible} with $R$ if, for each read $f\in
R$, there exists a path $p=\langle w_{1}, \ldots , w_{k}\rangle $ of $G$ such
that $r$ is a \emph{substring} of $s(w_{1})\cdots s(w_{k})$.
\end{definition}

\begin{problem}{Splicing Graph Reconstruction (SGR) Problem}
\label{pb:GF-construction}\\
Input:  a set $R$ of reads, all of length $l$, extracted from an unknown expressed gene $\ExGene$.
\noindent
Output: a splicing graph compatible with $R$.
\end{problem}

Clearly SGR can only be a preliminary version of the problem, as we are
actually interested into finding a splicing graph that is most similar to the
isoform graph associated to $\ExGene$.
Therefore we need to introduce some criteria to rank all splicing graphs compatible with $R$.
The parsimonious principle leads us to a natural objective function (albeit we
do not claim it is the only possibility): to minimize the
sum of the lengths of strings associated to the vertices (mimicking the search
for the shortest possible string coding region). In the rest of paper the SGR
problem will ask for a splicing graph minimizing such sum.

\section{Unique solutions to SGR}


In this section we will show some conditions must be satisfied if we want to
be
able to optimally solve the SGR problem.
Notice that, given an isoform graph $G_{I}$ there is a splicing graph $G_{S}$ naturally
associated to it, where the two graphs $G_{I}$ and $G_{S}$ are isomorphic
(except for the labeling) and the label of each vertex of $G_{S}$ is the
string of the corresponding vertex of $G_{I}$.

Let $R$ be an instance of SGR originating from an expressed gene
$\ExGene$. Then $R$ is a \emph{solvable} instance if:
(i) for each three blocks $b$, $b_1$ and $b_2$
\st $b$ and $b_1$ are consecutive in an isoform, $b$ and $b_2$ are
consecutive in another isoform, then $b_1$ and $b_2$ begin with
different characters. Also, for each three blocks $b$, $b_1$ and $b_2$
\st $b_1$ and $b$ are consecutive in an isoform, $b_2$ and $b$ are
consecutive in another isoform, then $b_1$ and $b_2$ end with
different characters;
(ii) for each subsequence $B_1$ of $B$, the
string $s(B_1)$ does not contain two identical substrings of length
$l/2$.
We will show here that our theoretical analysis can focus on solvable
instances, since
for each condition of solvable instance
we will show an example where there exists an optimal splicing graph --
different from the isoform graph -- compatible with the instance.

Condition (i).
Let $B=\{b, b_1, b_2\}$, and let $F=\{\langle b,b_{1} \rangle , \langle b,b_{2} \rangle \}$.
Moreover $s(b_{1})[1]=s(b_{2})[1]=x$, that is the strings of both blocks
$b_{1}$ and $b_{2}$ begins with the symbol $x$, which does not appear in any
other position of the string coding region.
Consider now the expressed gene $B'=\{b', b'_1, b'_2\}$, and
let $F'=\{\langle b',b'_{1} \rangle , \langle b',b'_{2} \rangle \}$, where
$s(b')=s(b)x$, $s(b'_{1})=s(b'_{1})[2:]$, and  $s(b'_{2})=s(b'_{2})[2:]$
(informally, the symbol $x$ is removed from $b_{1}$ and $b_{2}$ and attached
to the end of $b$).
It is immediate to notice that  $|s(b)|+|s(b_{1})|+|s(b_{2})| >
|s(b')|+|s(b'_{1})|+|s(b'_{2})|$ and
any set of reads that can be extracted from one gene can also be extracted
from the other.
A similar argument holds for the condition on the final character of the
string of each block.

Condition (ii).
Let us consider three blocks such that $b_{2} $ and $b_{3} $ contains the same
$l/2$-long substring $z$, that is $s(b_{2} )= p_{2}zs_{2}$ and  $s(b_{3} )= p_{3}zs_{3}$.
There are two isoforms: $\langle b_{1}, b_{2} \rangle $ and
$\langle b_{1}, b_{3} \rangle $.
Besides the isoform graph, there is another splicing graph
$v_{1}, $ $\ldots ,$ $v_{5}$ where $s(v_{1} )=s(b_{1} )p_{2}$, $s(v_{2} )=s(b_{1} )p_{3}$,
$s(v_{3})=z$, $s(v_{4} )=p_{2}$, $s(v_{5} )=p_{3}$
that is compatbile with any set of reads extracted from $\ExGene$.
The arcs of this splicing graphs are $(v_{1} , v_{2} )$ and $(v_{1}, v_{3})$.
Notice that the sum of  lenghts of the labels of this new splicing graph is
smaller than that of the isoform graph.

\section{Methods}
\label{sec:methods}

In order to investigate the  two main questions stated before, we propose
a method for solving the SGR problem. Our
approach to compute the isoform graph $G_S$ first  identifies the
vertex set $B_S$ and then the edge set $E_S$ of $G_S$.
Moreover we focus on fast and simple methods that can possibly scale to genome-wide data.
For ease of exposition, the discussion of the method assumes that
reads have no errors, and $l=64$.

The basic idea of our method is that we can find two disjoint subsets
$R_1$, and $R_2$ of the input set $R$ of reads, where reads of $R_1$, called
\emph{unspliced},
can be assembled to form the nodes in $B_S$, while reads of $R_2$,
called  \emph{spliced}, are an evidence of a
junction between two blocks (that is an arc of $G_S$).
We will  discuss  how our method deals with problems as
errors, low coverage.

The second main tenet of our algorithm is that each read is encoded by a
$128$-bit binary number, divided into a \emph{left fingerprint}  and a
\emph{right fingerprint} (respectively  the leftmost and the rightmost $64$
bits of the encoding). Then two reads $r_{1}$ and $r_{2}$  overlap for at
least $l/2=32$ base pairs iff the right fingerprint of $r_{1}$ is a substring of
the encoding of $r_{2}$ (a similar condition holds for the left
fingerprint of $r_{2}$).
Bit-level operations allows to look for such a substring  very quickly.

\begin{definition}
\label{def:spliced-unspliced}
Let $r$ be a read of $R$. Then $r$ is \emph{spliced} if there exists
another $r' \in R$, $s(r)\neq s(r')$, such that
$\prefix(r,k)=\prefix(r',k)$ or $\suffix(r,k)=\suffix(r',k)$, for $l/2
\leq k$. Moreover a read $r$ is \emph{perfectly spliced} if there
exists another $r' \in R$, $s(r)\neq s(r')$, such that the
longest common prefix (or suffix) of $r$ and $r'$ is exactly of length
$l/2$. A read that is not spliced is called \emph{unspliced}.
\end{definition}

In our framework, a junction site between two blocks $b_1$ and $b_3$,
that appear consecutively within an isoform, is detected when we find a
third block $b_2$ that, in some isoform, appears
immediately after $b_1$ or immediately before $b_3$. For illustrative
purposes, let us consider the case when $b_2$ appears immediately
after $b_1$ in an isoform (Figure~\ref{subfig:skipping}).  The strongest
possible signal of such junction
consists of two reads $r_1$ and $r_2$ such that
$ov(s(b_1),r_1)=ov(r_1,s(b_3)=l/2$ and $ov(s(b_2),r_2)=ov(r_2,s(b_3))=l/2$, that
is $r_1$ is cut into halves by the junction  separating $b_1$ and
$b_3$, while $r_2$ is cut into halves by the junction  separating
$b_2$ and $b_3$ (i.e. $r_1$ and $r_2$ are
perfectly spliced). In a less-than-ideal scenario, we still can find
two reads $r_1$ and $r_2$ sharing a common prefix (or suffix) that is
longer than $l/2$, in which case the two reads are spliced.

Notice that all reads extracted from the same block can be sorted so that any
two consecutive reads have large overlap.
More formally, we define a \emph{chain}  as a  sequence $C=\langle
r_1,r_2,\cdots,r_n \rangle$ of unspliced reads
where $ov(r_i, r_{i+1})= l/2$ for $1 \leq i < n$ (notice that the $\SH(r_{i})=\FH(r_{i+1})$, which
allows for a very fast computation).
Let $C$ be a chain.
Then the string of $C$ is the string $s(C)=\varphi(C)$, moreover $C$ is
\emph{maximal} if no supersequence of $C$ is also a chain. Under ideal conditions (i.e. no errors
and high coverage) $s(C)$ is exactly the string of a block.
Coherently with our reasoning, a perfectly spliced read $r$ is called a
\emph{link} for the pair  of chains $(C,C')$, if  $\FH(r)=\suffix(s(C),l/2)$ and
$\SH(r)=\prefix(s(C'),l/2)$. In this case we also say that $C$ and $C'$ are
respectively \emph{left-linked} and \emph{right-linked} by $r$.

Given  a set $R$ of reads extracted from the isoforms $F$ of an unknown
expressed region   $\ExGene$, our
algorithm outputs a likely  isoform graph
$G_R=(\mathcal{C},E_R)$, where $\mathcal{C}=\{C_1,\cdots,C_n\}$ is a
set of maximal chains that can be derived from $R$, and
$(C_i,C_j) \in E_R$ iff there exists
in $R$ a link for $(C_i, C_j)$.
The remainder of this section is devoted to show how we compute such
 graph efficiently even under less-than-ideal conditions.
The algorithm is organized into three steps that are detailed below.
In the first step we build a data structure to store the reads in
$R$. We use two hash tables which guarantee a fast access to the
input reads. The second step creates the nodes of
$G_R$ by composing the maximal chains of the unspliced reads of $R$.
In the last step of the creation of $G_R$, the maximal chains obtained
in the second step are linked.

\subsubsection{Step 1: Fingerprinting of RNA-Seq reads}

For each  read longer than $64$bp we extract some  substrings of $64$bp (usually
the leftmost and the rightmost) that are representative of the original read.
Then each $64$bp read is unambiguously encoded by a $128$-bit binary
number, exploiting the fact that we can encode each symbol with 2 bits as
follows: $\enc(a)=0=00_2$,
$\enc(c)=1=01_2$, $\enc(g)=2=10_2$, $\enc(t)=3=11_2$.
Since such encoding
is a one-to-one mapping between reads and numbers between $0$
and $2^{128}-1$, we will use interchangeably a string and
its fingerprint.
Moreover, given a read $r$, we define $\phi_1(r)$ (also called \emph{left
fingerprint}) and $\phi_2(r)$ (also called \emph{right fingerprint})
respectively as the leftmost and the rightmost $64$ bits of the encoding of $r$.

The main data structures are two tables $\Ll$ and $\Lr$,
both of which are indexed by $64$-bit fingerprints.
More precisely, $\Ll$ has an entry indexed by each left fingerprint,
while  $\Lr$ has an entry indexed by each right fingerprint.
The entry of $\Ll$, associated to the left fingerprint $f_{l}$,  consists
of a list of all the right fingerprints $f_{r}$ such that the concatenation
$f_{l}f_{r}$ is a read in the input set. The role of $\Lr$ is symmetrical.
The purpose of those tables is that they allow for a very fast labeling of
each read into unspliced or perfectly spliced reads.
In fact, a read $r$ is unspliced iff both the entry of $\Ll$ indexed by its left
fingerprint and the entry of \Lr indexed by its right
fingerprint are lists with only one element.
Moreover, let $f_{l}$ be a left fingerprint of some reads, let $f_{r,1}$
and  $f_{r,2}$ be two fingerprints in the list of \Ll indexed by $f_{l}$,
such that the first character of  $f_{r,1}$ is different from that of  $f_{r,2}$.
Then the two reads $f_{l}f_{r,1}$ and  $f_{l}f_{r,2}$ are perfectly spliced.
Also, constructing \Ll and \Lr requires time proportional to the number of the
input reads.

\subsubsection{Step 2: Building the set $\mathcal{C}$ of Maximal Chains}
The procedure $BuildChains$ described in
Algorithm~\ref{alg:build-chains} takes as input a set $R$ of RNA-Seq reads
and produces the set $\mathcal{C}$ of all  maximal chains
that can be obtained from $R$.
Let $R_1 \subseteq R$ be the set of the \emph{unspliced} reads.
The algorithm selects any read $r$ of $R_1$ and tries to find a \emph{right
extension} of $r$, that is another unspliced read $r_{r}$ such that $ov(r,r_r)= l/2$.
Afterwards the algorithm iteratively looks for a right extension of $r_{r}$,
until such a right extension no longer exists.
Then, the algorithm iteratively looks for a left extension of $r$, while it is possible.

Also, the time required by this procedure is proportional to the number of
unspliced reads. In fact, each unspliced read is considered only once, and
finding the left or
right extension of a read $r$ can be performed in constant time.
At the end, we will merge all  pairs of chains whose strings have an overlap
at least
$l/2$ bases long, or one is a substring of the other.
%
%
We recall that the maximal chains are the vertices of the isoform graph we
want to build.

\begin{algorithm}[h!]
\KwData{a set $R$ of RNA-Seq reads}
$\mathcal{C}\gets\emptyset$;
$R_1\gets \{r \in R | r$ is \emph{unspliced}$\}$\;
\While{$R_1\neq\emptyset$}{%
  $r\gets$ any read from $R_1$\;
  $R_1 \gets R_1 \setminus \{r\}$;
  $C\gets \langle r \rangle$;
  $r_1\gets r$\;
  \tcp{Extend the chain on the right}
  \While{$\exists$ a right extension $r_2 \in R_1$ of $r_{1}$}{%
        append $r_2$ to $C$\;
        $R_1 \gets R_1 \setminus \{r_2\}$;
    $r_1\gets r_2$\;
  }
  \tcp{Extend the chain  on the left}
  \While{$\exists$ a left extension $r_2 \in R_1$ of $r$}{%
    prepend $r_2$ to $C$\;
    $R_1 \gets R_1 \setminus \{r_2\}$;
    $r\gets r_2$\;
  }
  $\mathcal{C} \gets \mathcal{C} \cup {C}$\;
}
\Return{$\mathcal{C}$};
\caption{BuildChains($R$)}
\label{alg:build-chains}
\end{algorithm}

\subsubsection{Step 3: Linking Maximal Chains}
Our algorithm computes the  arcs
of the output graph using the set $R_2$ of \emph{perfectly spliced} reads
and the set $\mathcal{C}$ of maximal chains computed in the previous step.
More precisely, given a perfectly spliced read $r$, we denote with
$\mathcal{D}(r)$ and $\mathcal{A}(r)$ the set of maximal chains
that are, respectively, left-linked
and right-linked by $r$.
Moreover each such pair will be an arc of the graph.

\subsection{Isomorphism between predicted and true isoform graph}
\label{sec:validation}

Let $R$ be an instance of SGR originating from an expressed region
$\ExGene$.
We can prove that a simple polynomial time variant of our method computes a
splicing graph  $G_R$  that is isomorphic to the true isoform graph, when $R$
is a good instance.
More precisely,  $R$ is a \emph{good} instance if it is solvable and
(iii) for each
isoform $f$ there exists a sequence $r_{1}, \ldots ,  r_{k}$ of
reads such that each position of $f$ is covered by some read in $r_{1}, \ldots
, r_{k}$ (i.e. $s(f)$ is equal to the fusion of $r_{1}, \ldots , r_{k}$) and
$|ov(r_{i}, r_{i+1})|\ge l/2$ for
each $i<k$.

First of all, notice that two reads $r_{1}$ and $r_{2}$ with overlap at least
$l/2$ can be extracted from the same isoform.
Let us build a graph $G$ whose vertices are the reads and an arc goes from
$r_{1}$ to $r_{2}$ if and only if $ov(r_{1}, r_{2})\ge l/2$ and there exists
no read $r_{3}$ such that $ov(r_{1}, r_{3})\ge ov(r_{1}, r_{2})$ and $ov(r_{3}, r_{2})\ge l/2$.
By the above observation and by condition (iii) there is a 1-to-1 mapping
between maximal paths in such graph and isoforms and the string of an  isoform
is equal to the fusion of the reads of the corresponding path.
Compute the set $R_{1}$ of all $l$-mers that are substrings of the string of
some isoforms.
Then classify all reads of $R_{1}$ into unspliced, perfectly spliced and
(non-perfectly) spliced reads, just as in our method.
Notice that the halves of each perfectly spliced read are the start or the end
of a block.
Assemble all unspliced reads into chains where two consecutive reads have
overlap $l-1$ (each unspliced read belongs to exactly one chain), using the
putative start/end $l/2$-mers computed in the previous step to trim the
sequences of each block.
At the end, each perfectly spliced read links two blocks of the splicing graph.
We omit the proof that this algorithm computes the isoform graph.

\subsection{Low coverage, errors and SNP detection}

We will consider here what happens when the input instance does not satisfy
the requirements of a good instance.
There are at least two different situations that we will have to tackle:
data errors and insufficient coverage.

One of the effects of the chain merging phase is that most errors are corrected.
In fact the typical effect of a single-base error in a read $r$ is the
misclassification of $r$ as spliced instead of unspliced, shortening or
splitting some chains.
Anyway, as long as there are only a few errors, there exist some overlapping
error-free unspliced reads that span the same block as the erroneous read.
Those unspliced reads allow for the correction of the error and the
construction of the correct chain spanning the block.

Notice that the chain merging also lessens the impact of partial coverage --
that is when we do not have all possible $l$-mers of a block.
When working under full coverage, we can identify a sequence of reads
spanning a block and such that two consecutive reads have overlap equal to
$l-1$, while the chain merging step successfully reconstruct the blocks with
reads overlapping with at least $l/2$ characters.

A similar idea is applied to pairs of reads $r_{1}$, $r_{2}$ with $ov(r_{1},
r_{2})\ge l/2$ and that are likely to span more than one block. Those reads
can be detected by analyzing the hash tables.
In this case a set of additional reads, compatible with the fusion of $r_{1}$
and $r_{2}$ is added to the input set, obtaining an enriched set which
includes the perfectly spliced reads required to correctly infer the junction,
even when the original reads have low coverage.

Also the fact that the definition of perfectly spliced read asks for two reads
with the same left (or right) fingerprint, makes our approach more resistant
to errors, as a single error is not sufficient to generate an arc in the
splicing graph.

Finally, we point out that our approach allows for SNP detection.
The main problem is being able to distinguish between errors and SNPs: let us
consider an example that illustrates a strategy for overcoming this problem.
Let $e$ be an exon containing a SNP, that is $s(e)$ can be $yaz$ or $ybz$,
where $y$ and $z$ are two strings and $a$, $b$ are two characters.
Moreover notice that, since this situation is a SNP, roughly the same number
of reads support  $yaz$ as $ybz$.
Therefore, as an effect of our read enrichment step, there are two reads $r_{1}$ and $r_{2}$ \st $r_{1}$ supports
$yaz$ and $r_{2}$ supports $ybz$, and $\FH(r_{1})=\FH(r_{2})$ or  $\SH(r_{1})=\SH(r_{2})$.
Equivalently, $r_{1}$ and $r_{2}$ are two spliced reads supporting the SNP.
This case can be easily and quickly found examining the list of reads
sharing the left (or right) fingerprints and then looking for a set of reads
supporting the SNP (again exploiting the fact that the fingerprint of a half
of the reads in the set is known).

\subsection{Repeated sequences: stability of graph $G_R$}

When  compared with approaches  based on  de Bruijn graphs, our method is
stable \wrt repeated sequences shorter than $l/2$, that is our method is not
negatively influenced by those short repeats.
Let us state formally this property. An algorithm to solve the SGR problem
is \emph{$k$-stable} if and only if we obtain a new isoform set $F'$ from $F$
by disambiguating each $k$-long substring that appears in more than one
isoform but originates from different parts of the string coding region, then the graph obtained
from $F$ is isomorphic to that obtained from $F'$.
Notice that de Brujin graphs are highly sensitive to this case, as they must
merge $k$-long repetitions into a single vertex.
On the other hand, our approach can avoid merging $(l/2-1)$-long repetitions,
as the chain construction step is based on finding $l/2$-long identical
substrings in the input reads.
Validating this property is one of the features of our experimental analysis.

Let us consider the following example.
Let $s_{A}= yvz$ and $s_{B}=u v w$ be respectively a block of gene   $A$ and $B$,
therefore $s_{A}$ and $s_{A}$ belong to two different weakly connected
components of the isoform graph.
Assume that $v$ is a sequence of length $k < l/2$, where $k$ is the parameter
length used in the construction of de Brujin graphs
(\ie the vertices correspond to $k$-mers),     and consider the case
where reads coming from both genes are given in input to compute a splicing graph.
If the instance is good, our approach is able to reconstruct the isoform
graph, while a typical algorithm based on de Brujin graphs would have a single
vertex for $x$.
Notice also that the resulting graph would be acyclic, hence the commonly used
technique of detecting cycles in the graph to determine if there are
duplicated strings is not useful.

\section{Experimental Results}
\label{sec:results}

We have run our experimental analysis on
simulated (and error-free) RNA-Seq data
obtained
from the transcript isoforms annotated for a subset of $112$ genes extracted
from the $13$ ENCODE 
regions used as training set in the EGASP competition (we refer the interested
reader to~\cite{Guigo2006} for the complete list of regions and
genes). We have chosen those genes because they are well
annotated and, at the same time, are considered quite hard to be analyzed by tools
aiming to compute a gene structure, mainly due to the presence of repeated regions.
Moreover, the presence of high similar genes makes this set very hard to be analyzed as a whole.
Also, we decided to focus on a relatively small number of (hard to analyze)
genes so that we could manually inspect the results to determine the causes of
incorrect predictions.

A primary goal of our implementation is to use only a limited amount of
memory, since this is the main problem with currently available programs~\cite{Grabherr2011}.
In fact, we have run our program on a workstation with two quad-core
Intel Xeon 2.8GHz processors and 12GB RAM.
Even on the largest dataset, our program has never used more $30$ seconds or
more than $250$MB of memory. Our implementation is available under AGPLv3 at \url{http://algolab.github.com/RNA-Seq-Graph/}.

Now let us detail the experiments. For  each of the $112$ ENCODE
genes, the set of  the annotated full-length transcripts has been
retrieved from NCBI GenBank.
From those transcripts we have extracted two sets of $64$bp substrings
corresponding to our simulated reads.
The first set consists of all possibile $64$-mers and corresponds to the
maximum possible coverage (\emph{full coverage} dataset), while the second set
consists of a random set of $64$-mers simulating an $8$x coverage
(\emph{low coverage} dataset).
In the first experiment we have analyzed the behavior on the full coverage
dataset where data originating from each gene have been elaborated separately
and independently.
The second experiment is identical to the first, but on the low coverage dataset.
Finally, the third experiment has been run on the whole full coverage dataset,
that is all reads have been elaborated together and without any indication of
the gene they were originating from.
Notice that the whole full coverage dataset consists of  $1.4$ Million unique $64$bp simulated reads.
For each input gene, the true isoform graph has been reconstructed from the
annotation.

To evaluate how much the splicing graph computed is similar to the isoform
graph, we have designed a general procedure to compare two labeled graphs,
exploiting not only the topology of the graph, but also the labeling of each
vertex with the goal of not penalizing small differences in the labels (which
corresponds to a correct detection of the AS events and a mostly irrelevant
error in computing the nucleotide sequence of a block).
Due to space constraints, we omit the details of the graph comparison procedure.

In all experiments, the accuracy of our method is evaluated by two standard measures,
\emph{Sensitivity} (Sn) and \emph{Positive Predictive Value}
(PPV) considered at vertex and arc level.  Sensitivity is
defined as the proportion of vertices (or arcs) of the isoform graph that have
been correctly predicted by a vertex (or arc) of the computed splicing graph,
while PPV is
the proportion of the vertices (or arcs) of the splicing graph that correctly
predict a vertex (or an arc) of the isoform graph.

The goal of the first experiment (each gene separately, full coverage) is to
show the soundness of our approach, since obtaining satisfying results under
full coverage is a requisite even for a prototype.
More precisely our implementation has perfectly reconstructed the isoform
graph of $43$ genes (out of $112$), that is Sn and PPV are $1$ both at vertex
and arc level. Notice that the input instances are usually not good instances,
mostly due to the presence of short blocks or relatively long repeated regions,
therefore we have no guarantee of being able to reconstruct the isoform graph.
Moreover we have obtained average  Sn and PPV values that are $0.86$
and $0.92$ at vertex level, respectively, and $0.72$
and $0.82$ at arc level, respectively.
Also, the median values of Sn and PPV are $0.91$ and $1$ at vertex level,
$0.86$ and $0.98$ at arc level, respectively.

The second experiment (separated gene, $8$x coverage) is to
study our method under a more realistic coverage.
In this case, we have perfectly reconstructed the isoform
graph of $39$ genes (out of $112$), and we have obtained average
Sn and PPV values that are respectively $0.87$, $0.91$ at vertex level and
$0.75$, $0.81$ at arc level. The median values of Sn and PPV are
$0.93$ and $1$ at vertex level,
$0.84$ and $0.91$ at arc level, respectively.

The main goal of the third experiment (\emph{whole dataset}, full coverage) is
to start the study of the scalability of our approach towards a genome-wide
scale, determining if repetitions that occur in different genes are too high
obstacles for our approach. A secondary goal is to determine if our
implementation is stable, that is the computed splicing graph is not too
different from the disjoint union of those computed in the first experiment.
In this experiment the expected output of  is a large
isoform graph $\GS$ with $1521$ vertices and $1966$ arcs, with a 1-to-1
mapping between
connected components and input genes.
To determine such mapping, we have used
a strategy based on BLAST~\cite{blast1}, aligning labels of the vertices and
the genomic sequences.
Due to space constraints, we omit the
description of such strategy, but we report that only $7$ connected component
have been  mapped to more than one gene --
$4$ of them are genes with very similar sequence composition (i.e. CTAG1A, CTAG1B
and CTAG2).

\begin{figure}[bt]
  \centering
  \begin{minipage}[t]{4cm}
\includegraphics[width=3cm]{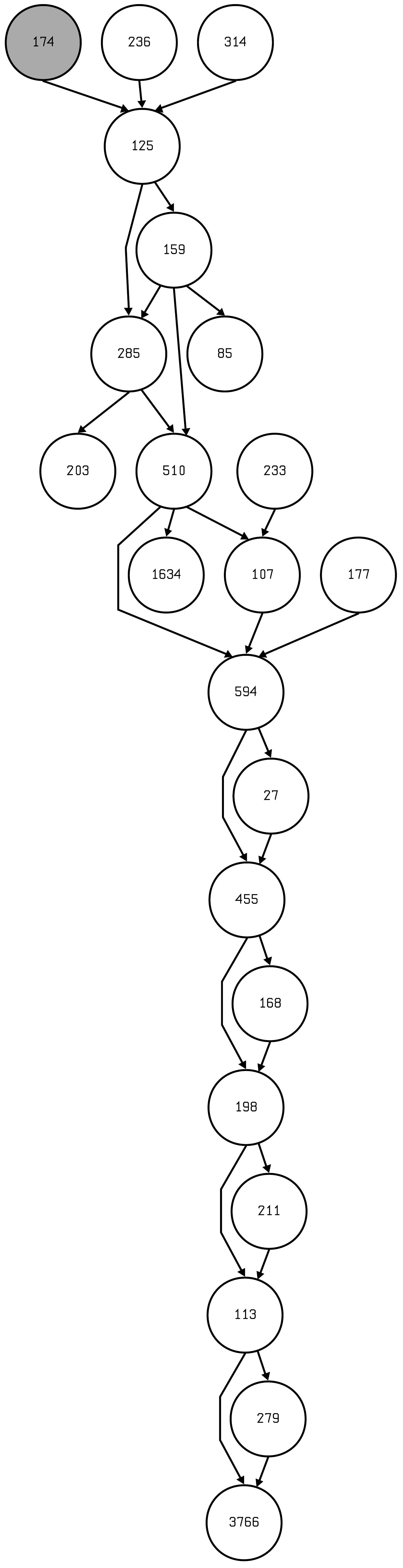}
\end{minipage}
\begin{minipage}[t]{4cm}
\includegraphics[width=3cm]{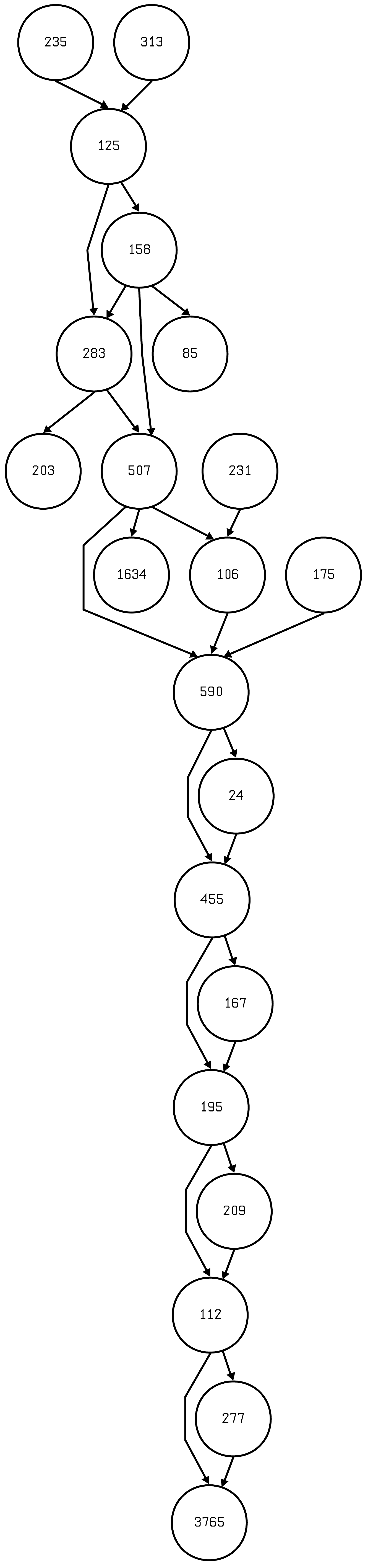}
\end{minipage}
\caption{Gene POGZ. The isoform graph  (on the left)
  and the splicing graph (on the right) predicted in the third experiment.
  The difference between the
  two graphs consists of the gray vertex that is missing in the splicing graph.
}
\label{fig:POGZ}
\end{figure}

In practice, such a 1-to-1 mapping exists for almost all components.
Moreover only $17$ genes have been associated to more than one connected
components.
Overall results are quite similar to those of the first experiment.
In fact,  the number of correctly
identified vertices goes from $1303$ (first experiment) to
$1274$ (third experiment).
Similarly, the number of correctly
identified arcs goes from $1415$ to $1396$ -- the quality of our
predictions is only barely influenced by the fact that the program is run on
the data coming from $112$ different genes.
The overall vertex sensitivity is $0.837$, the vertex positive
predicted value is $0.778$, while the arc sensitivity is $0.71$ and the
arc positive predicted value is $0.679$.
Figure~\ref{fig:POGZ} shows the isoform graph and the predicted graph
for the gene POGZ.

The final part of our analysis is a comparison with
Trinity~\cite{Grabherr2011} -- the most advanced available tool for full-length
transcript reconstruction from RNA-Seqs without a reference genome, to
determine how much it is stable.
We have run Trinity on the two full coverage datasets, corresponding to the
first and third experiments.
Since Trinity computes transcripts and not the splicing graph, we use the
variation of number of predicted full-lengths transcripts as a proxy for the
(in)stability of the method.
We observed that, for the two datasets, Trinity has predicted $2689$ and
$1694$ full-length transcripts (on the genes from which the simulated read are
generated, there are $1020$ annotated
transcripts).
The variation is significant and hints at a desired property of our algorithm
that is not shared with other state-of-the-art tools.

\begin{figure}[tp!]
\centering
\includegraphics[width=5cm]{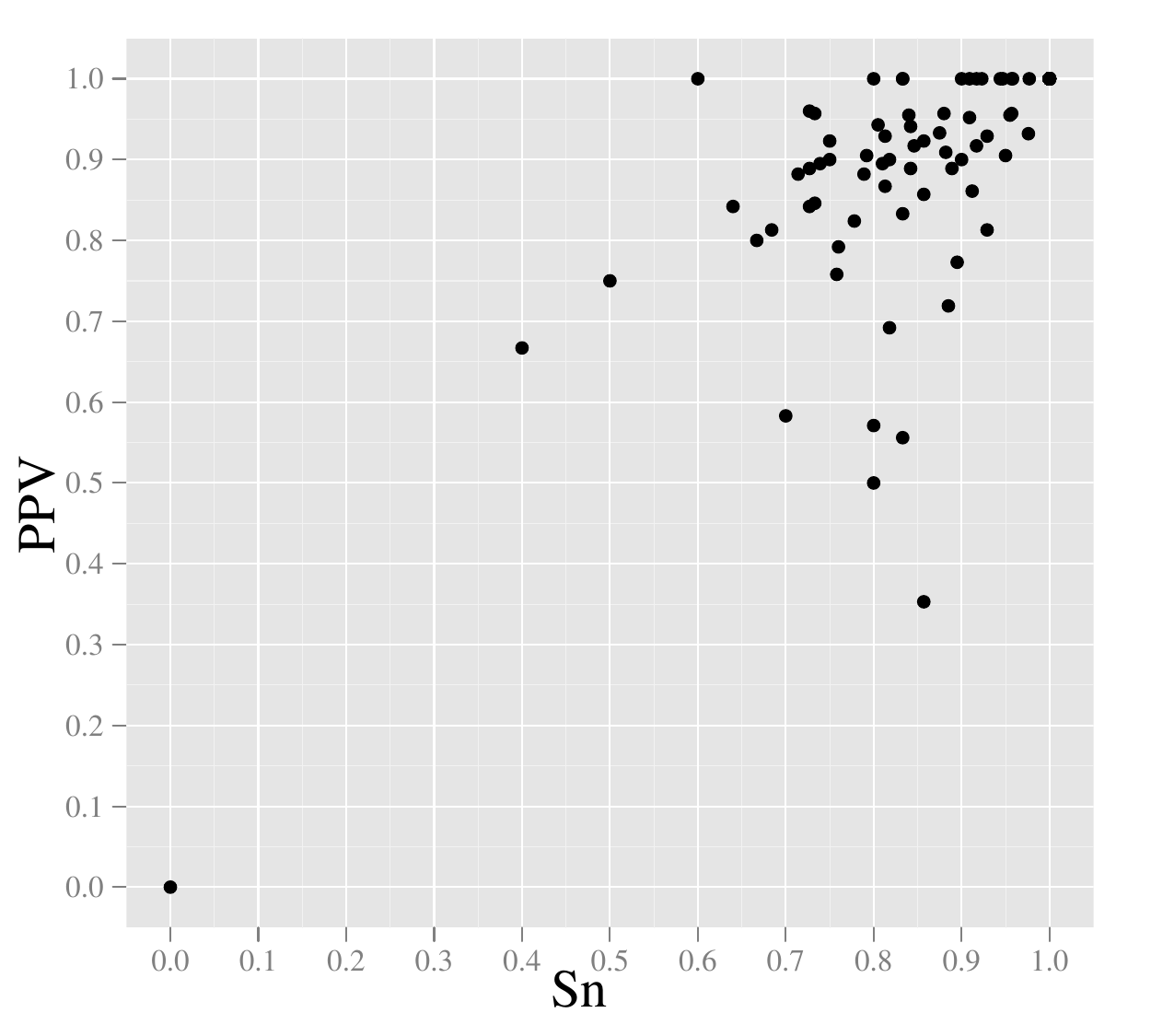}
\caption{Second experiment: Sn and PPV values at vertex level.
Each point represents a gene.}
\end{figure}


\begin{figure}[tp!]
\centering
\includegraphics[width=5cm]{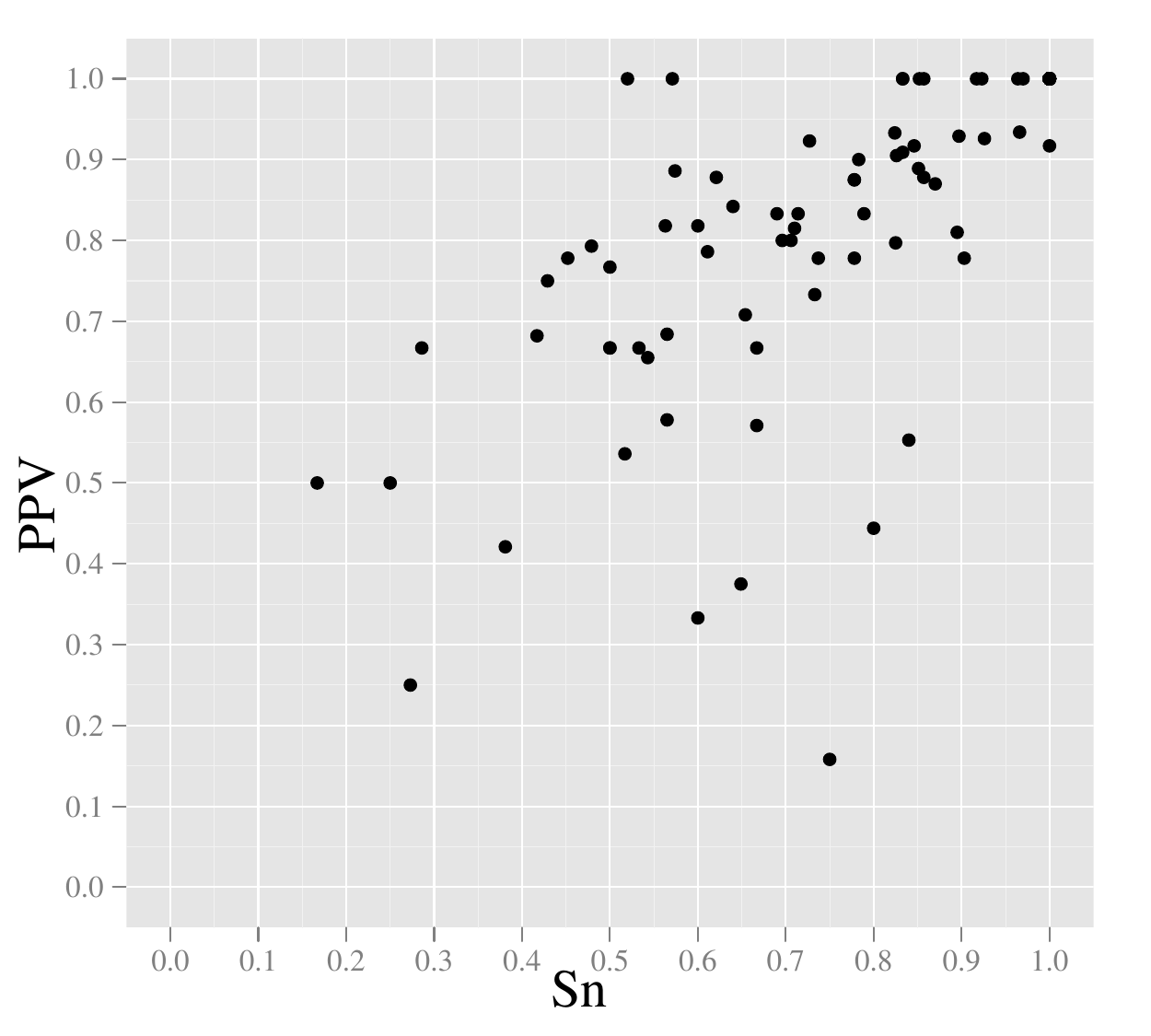}
\caption{Second experiment: Sn and PPV values at arc level.
Each point represents a gene.}
\end{figure}



\end{document}